\documentclass[12pt]{emulateapj}
\usepackage{epsfig}
\shorttitle{IR Counterpart to GC X-ray Source}
\shortauthors{Mikles et al.}

\begin{document}
\title{Identification of the Infrared Counterpart to a Newly Discovered X-ray Source in the Galactic Center}
\author{Valerie J. Mikles}
\affil{Department of Astronomy, University of Florida, Gainesville, FL 32611}
\email{mikles@astro.ufl.edu}

\author{Stephen S. Eikenberry}
\affil{Department of Astronomy, University of Florida, Gainesville, FL 32611}
\email{eikenberry@astro.ufl.edu}

\author{Michael P. Muno\altaffilmark{1}}
\affil{University of California, Los Angeles}
\altaffiltext{1}{Hubble Fellow}
\email{mmuno@astro.ucla.edu}

\author{Reba M. Bandyopadhyay}
\affil{Department of Astronomy, University of Florida, Gainesville, FL 32611}
\email{reba@astro.ufl.edu}

\and 

\author{Shannon Patel}
\affil{University of California, Santa Cruz}
\email{patel@astro.ucsc.edu}

\begin{abstract}
We present first results of a campaign to find and identify new compact objects in the Galactic Center. Selecting candidates from a combination of Chandra and 2MASS survey data, we search for accretion disk signatures via infrared spectroscopy. We have found the infrared counterpart to the {\it Chandra} source CXO J174536.1-285638, the spectrum of which has strong Br$\gamma$ and HeI emission. The presence of CIII, NIII, and HeII indicate a binary system. We suspect that the system is some form of high-mass binary system, either a high-mass X-ray binary or a colliding wind binary.
\end{abstract}


\section{Introduction}

Heavily obscured by dust, the Galactic Center (GC) is virtually unobservable at visible wavelengths. However, infrared (IR) observations reveal a dense stellar population traced by X-ray sources, including compact objects, massive star clusters, and a supermassive black hole. The GC has been known to be home to X-ray emitting compact objects since the beginning of such studies in the 1960s \citep[see, e.g.,][]{giac72,gur72,for78}. Identifying and studying the population of compact objects in the GC gives important insights into the massive star formation history of this region. With the sub-arcsecond resolution of modern IR and X-ray instruments, we are able to penetrate the extinguishing veil of the interstellar medium and explore this exciting region.

In this paper, we present the first results of a campaign to locate and identify compact objects in the GC. Recent {\it Chandra X-ray Observatory} observations of the GC revealed a new population of faint X-ray sources with $L_X \sim 10^{31} - 10^{33}$ erg~s$^{-1}$ \citep{mun04}. We cross-correlate these {\it Chandra} observations with the archived 2MASS catalog, identifying possible IR counterparts to the X-ray sources. We have observed nine sources at the Infrared Telescope Facility (IRTF) and in this paper announce the first definitive identification of an IR counterpart to one of these new {\it Chandra} sources: CXO J174536.1-285638 (internal catalog ID, Edd-1). In Section 2, we discuss our observations and analysis. In Section 3, we compare Edd-1 to other X-ray emitting systems that contain high-mass stars including high-mass X-ray binaries (HMXBs) and colliding-wind binaries (CWBs). In Section 4, we summarize our conclusions.

\section{Observations and Analysis}

We use archival {\it Chandra} observations of the GC region to identify $\sim$400 serendipitous X-ray sources within $\sim$10-arcmin of Sgr A* (equivalent to a projected size of $\sim 23$ pc at a distance of 8~kpc \citep{mcn00}) with $<$1-arcsec positional accuracy. We
cross-correlate this sample with the 2MASS survey in the $K_S$ band, breaking them into categories of possible
matches ($<$1.5-arcsec positional difference between {\it Chandra} and nearest 2MASS source) and close non-matches (1.5-3.0-arcsec difference). This produces a list of $\sim$180 {\it Chandra} sources with possible IR counterparts. We then use the close non-matches
to estimate the false positive rate for association in each {\it Chandra} observation by using the surface density of the close non-matches to estimate the probability of obtaining a random match. From this probability, we eliminate all regions for which we expect $>$35\% of the possible
matches to be random coincidences. We then compare the ratios of the X-ray to IR flux for the matched sources, and eliminate those with low ratios ($log(F_X/F_{IR}) < -2$, reddened) as being likely due to stellar atmospheric emission rather than compact objects. Finally,
we use the J-K$_S$ colors of the potential 2MASS counterparts to remove all candidates which are too blue (J-K$_S <$2.0 mag)
to be located in the reddened GC. We choose a criterion $A_V \approx 12$ in order to ensure that there was little chance of any star physically located in the GC being excluded from our list of candidates. We thus produce a list of X-ray sources with probable IR counterparts at or beyond the distance of the GC.
Using this astrometric color criteria, we identify Edd-1 as a potential compact object. Follow-up IR spectroscopy with IRTF has revealed a source rich in emission features signifying the presence of a hard radiation field consistent with an associated IR and X-ray source \citep[see, e.g.,][]{clark00, var04}. In this section, we discuss the IR and X-ray observations of Edd-1.

\subsection{Infrared}
According to the 2MASS catalog, Edd-1 has IR magnitudes of: J=15.56 $\pm$ 0.08, H=12.11 $\pm$ 0.06, and K$_S$ =10.33 $\pm$ 0.07. 
On 2005 July 1 UT we obtained J, H, and K band (1.1-2.4 $\mu$m) spectra of Edd-1 using SpeX on IRTF \citep{rayner03}. Nodding along the length of the slit, we obtained six 120s exposures for a total exposure time of 720s. In the short-wavelength, cross-dispersed mode, we attained a resolution of $R \sim 1200$ over the JHK bandpass. The resolution estimation is confirmed by measurements of OH sky lines. Target observations were followed by observations of the G0V star HR~6836 at similar airmass for removal of telluric absorption features. Using the standard SpeX macro cal\_sxd\_0.5, we obtained flat fields and wavelength calibration. 

We extract spectra using the standard SpexTool procedure for AB nodded data, resulting in a series of sky-subtracted, wavelength-calibrated spectra \citep{vacca03,cushing04}. Hot pixels, cosmic rays, and other non-intrinsic spectral features are removed using the IDL code ``xcleanspec'' and then individual spectra are summed using ``xcombspec''. Both of these programs are included in the SpexTool package. We interpolate over the intrinsic Brackett absorption features in the G0V-star spectrum, then divide the target spectrum by the G0V-star in order to remove atmospheric absorption bands. We multiply the resultant spectrum by a 5900~K blackbody spectrum, corresponding to the temperature of the G0V-star. 

\begin{figure*}
\epsscale{0.7}
\plotone{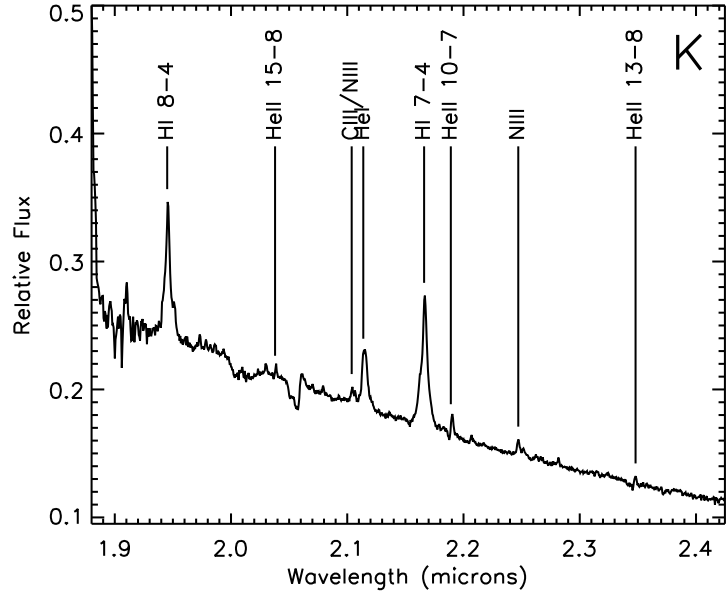}
\caption{The K-band spectrum of Edd-1 shows strong Br$\gamma$, Br$\delta$, and HeI emission. P~Cygni profiles are seen in several of the Helium lines suggesting a Helium wind around a massive star. }
\label{fig1}
\end{figure*}

\begin{figure*}
\epsscale{0.7}
\plotone{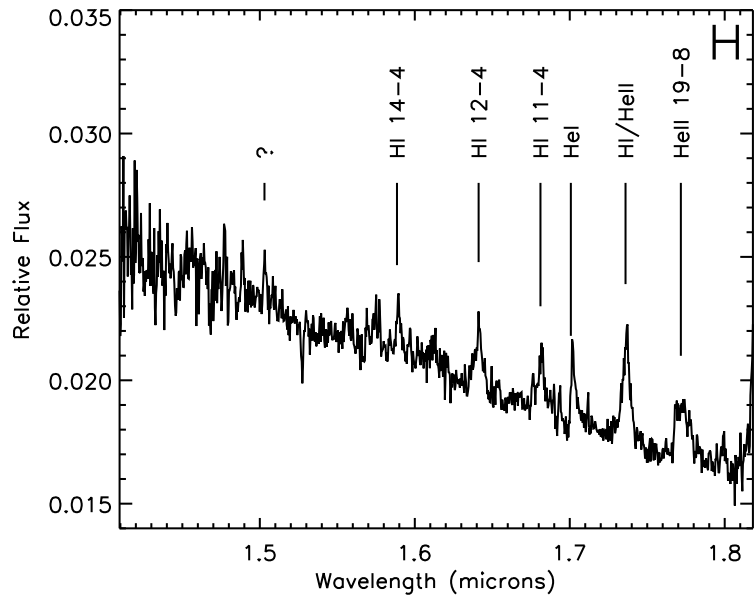}
\caption{Brackett series emission dominates the H-band spectrum of Edd-1.}
\label{fig2}
\end{figure*}

\begin{figure*}
\epsscale{0.7}
\plotone{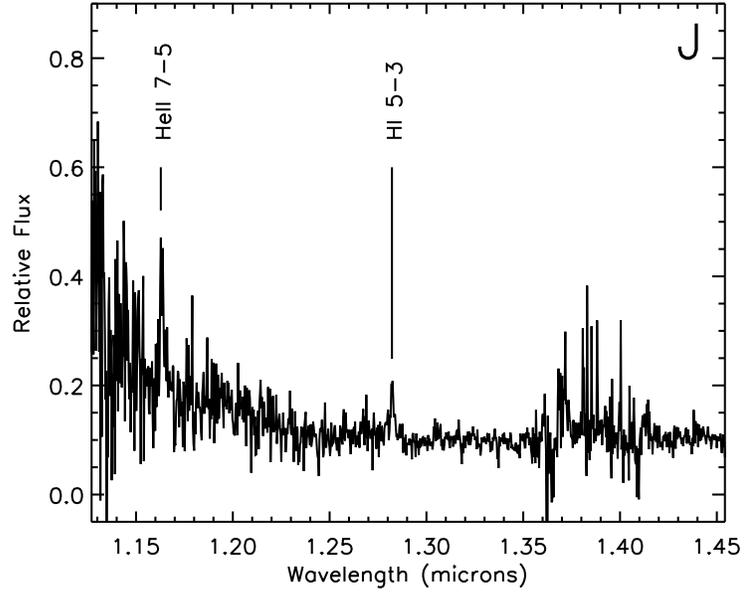}
\caption{The $Pa\beta$ line is clearly visible in the J-band spectrum of Edd-1. Atmospheric noise distorts the spectrum at $\lambda =1.35-1.42 \mu$m and at $\lambda < 1.15$}
\label{fig3}
\end{figure*}

\begin{figure*}
\plotone{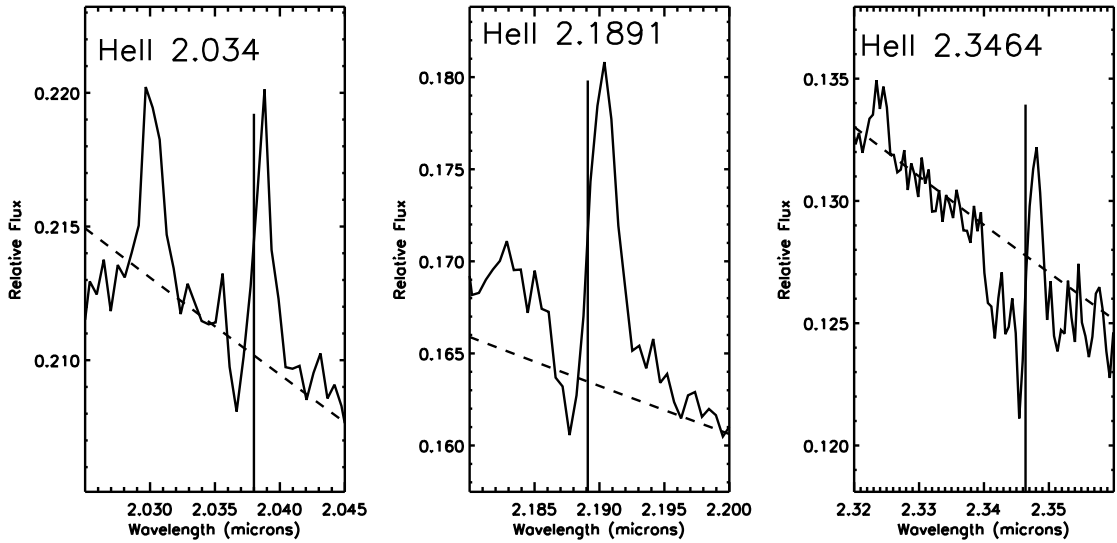}
\caption{P Cygni profiles for the Helium lines. The dotted line shows the approximate continuum level. The vertical line is placed at the vacuum center wavelength.}
\label{fig4}
\end{figure*}

To estimate the reddening toward Edd-1, we assume a GC distance of 8kpc \citep{mcn00}. We estimate the infrared extinction toward Edd-1 in two ways. First,  we assume a \citet{cardelli89} reddening law and an intrinsic $(H-K)_0 =0$ for a hot star/disk, which gives a value of $A_V=29$ mag. As a second method, we estimate the reddening by fitting the K-band spectral continuum to the Rayleigh-Jeans tail of a $T > 10^4 K$ blackbody. The fit corresponds to $A_V = 33$ mag. While both values are typical of the GC, we adopt the more conservative value of $A_V$= 29 mag in this paper. Given this, the dereddened magnitudes are: J=7.3, H=6.9, and K$_S$ =6.9.
\begin{deluxetable*}{llllll}
\tablecaption{{\bf IDENTIFIED LINES}}
\tablewidth{0pt}
\tablehead{\colhead{Band} & \colhead{Line}  & \colhead{$\lambda _c$ ($\mu$m)} & \colhead{EW ($\AA$) } & \colhead{$V_{FW}$} & \colhead{Comments} } 
\startdata
\tableline
\tableline
J & HeII 7-5             & 1.163 & $-29 \pm 4$        & 450            &        \\
\tableline		 													  
J & HI 5-3 (Pa$\beta$)              & 1.282 & $-19 \pm 2$        & 310	        & blend?       \\
\tableline		 													  
\tableline		 													  
H & unknown              & 1.503 & $-1.9 \pm 0.4$     & 340		&        \\
\tableline		 													  
H & HI 14-4 (Br14)             & 1.589 & $-2.3 \pm 0.3$     & 280		&        \\
\tableline		 													  
H & HI 12-4 (Br12)              & 1.641 & $-6.6 \pm 0.4$     & 470		&        \\
\tableline		 													  
H & HI 11-4  (Br11)             & 1.681 & $-5.9 \pm 0.6$  & 570	&        \\
\tableline		 													  
H & HeI 4D-3P, 3D-3$P_0$ & 1.701 & $-2.7 \pm 0.4$     & 370		&        \\
\tableline		 													  
H &  HI 10-4, HeII 20-8   &  1.736    &   bl        & 	bl	& blend       \\
\tableline		 													  
H & HeII 19-8            &  1.772     & $-10.0 \pm0.5$     &	2160	&        \\
\tableline		 													  
\tableline		 													  
K & HI 8-4  (Br$\delta$)              & 1.945 & $-36.2 \pm 0.6 $     & 640	& HeI blend?      \\
\tableline		 													  
K & CIII/ NIII           & 2.104 & bl    & bl		& blend       \\
\tableline		 													  
K & HeI 4S-3P, 1S-1$P_0$ & 2.114 & $-13.8 \pm 0.5$     & 590		& blend? CIII/NIII      \\
\tableline																  
K & HI 7-4 (Br$\gamma$)               & 2.166 & $-36.6 \pm 0.3$     & 710	&        \\
\tableline															  
K & NIII                 & 2.247 & $-1.7 \pm 0.2$     & 310		&        \\
\tableline																  
\enddata
\tablecomments{The identified lines transitions, vacuum wavelength, equivalent width, and full-width velocity. The full-width velocity has been corrected for the intrinsic line width of the instrument. Vacuum line centers are obtained from \citet{schu05, morris96, wallace00, hanson96, figer97}. Uncertainty in velocity is up to 70km/s due to uncertainties in measured line width.}
\label{tbl-1}
\end{deluxetable*}
Figures \ref{fig1}, \ref{fig2}, and \ref{fig3} show the spectra for the K, H, and J bands respectively, dereddened by  $A_V$= 29 mag. The spectra are dominated by strong hydrogen emission lines, including Paschen-$\beta$, Brackett-$\gamma$, and Brackett series lines Br10 - Br14. The Br13 line is not distinguishable in our spectrum. We observe two neutral Helium lines ($\lambda$~1.701 and 2.113~$\mu$m) and six HeII transitions ($\lambda$~1.163, 1.736, 1.772, 2.189, 2.038, and 2.348~$\mu$m). The HeII $\lambda$~1.736 is blended with the Br10 line. In the K-band we also observe metal lines from CIII and NIII, consistent with an accretion signature or a colliding wind system. We fit a Gaussian function to the line profile to determine the line centers and FWHM. We estimate the spectral resolution of the instrument by measuring the width of OH sky lines and correct our measured line widths accordingly. We give the line centers, equivalent widths, and full-width velocity in Table \ref{tbl-1}. Most of the emission lines are broad with a full-width velocity above 300 km/s. The Br$\gamma$ lines is strongest and has a full-width velocity of 710km/s. Given our resolution, it is not trivial to deconvolve the Br$\gamma$ line from neighboring HeI emission at $\lambda 2.162-2.166 \mu m$. Detailed modeling is required to accurately assess the Br$\gamma$ line equivalent width independently of the HeI contribution and such analysis is beyond the scope of this paper. Both \citet{morris96} and \citet{hanson96} study the IR spectra of massive stars without quantifying the relative contributions. \citet{morris96} note that the apparent asymmetry of the Br$\gamma$ line is likely caused by HeI contribution, but says the contribution is relatively weak in the context of LBV and Ofpe/WN9 stars. Because our speculations on spectral classification are based on broad X-ray and IR spectral features and not on specific line ratios, it is unlikely that the composite nature of the line affects our results here. Three HeII lines in Edd-1 show P~Cygni profiles at 2.034 $\mu$m, 2.189 $\mu$m, and 2.348 $\mu$m (Fig. \ref{fig4}, Table \ref{tbl-2}). We calculate the differential velocity from the line center to the blue edge and get an average $v =170$ km/s. Error due to pixel size and peak location is ~70km/s.

\begin{deluxetable}{llll}
\tablecaption{{\bf P~Cygni Line Velocity}}
\tablewidth{0pt}
\tablehead{\colhead{Band} & \colhead{Line}  & \colhead{$\lambda _{vac}$ ($\mu$m)} & \colhead{$v_{blue}$ (km/s)}  }
\startdata
K & HeII 15-8            & 2.0379 & 180     \\
\tableline
K & HeII 10-7            & 2.1891 & 150     \\
\tableline
K & HeII 13-8            & 2.3464 & 180 \\
\tableline
\enddata
\tablecomments{The velocity is calculated based on the difference between the blue edge and vaccum central wavelength. Line centers are as referenced in Table \ref{tbl-1}. Because of discrepencies in the location of the HeII 13-8 line, we estimate the central wavelength using the equation $0.0911138 Z^{-2} n_i^2n_j^2/(n_j^2- n_i^2)$. Uncertainties due to pixel size and peak location give a $\Delta V = 70km/s$. }
\label{tbl-2}
\end{deluxetable}

\subsection{X-ray}

\citet{mun04b} examined the spectrum and variability of the X-ray emission from 
Edd-1 as part of a study of 
$\approx$2000 X-ray sources detected toward the Galactic Center. Here, we summarize the properties of the X-ray 
source derived from that study, based on 626 ks of {\it Chandra} observations 
taken between 1999 September and 2002 June. The analysis is described
in detail in \citet{mun04}.
First, the pulse heights of each event were corrected to account for
position-dependent charge-transfer inefficiency \citep[CTI;][]{tow02},
and the lists were cleaned using standard tools in CIAO version 3.2
to remove those that did not pass the standard ASCA grade filters, 
that did not fall within the good time intervals defined by the 
Chandra X-ray center, or that occurred during intervals when 
background rate flared to $\ge 3\sigma$ above the mean level. 
We then extracted counts from a contour enclosing 90\% of the point
spread function around Edd-1, and binned them 
as a function of time to create light curves, and of energy to create 
spectra. The background was estimated from an annular region surrounding
the source. For each observation, we obtained the instrumental response 
functions from \citet{tow02} and computed effective area functions using 
the CIAO tool {\tt mkarf}, and averaged these, weighted by the numbers 
of counts in each observation.

The X-ray emission in Edd-1 varied in 
intensity by a factor of $\sim$3 in the 2-8keV range. The three observations in which 
the source was faint lasted only $\sim$30 ks in total, which is a
small fraction of the total exposure. They did not provide enough signal
to test whether the spectrum varied along with the flux, so we only
examined the average spectrum in detail.

\begin{figure}
\plotone{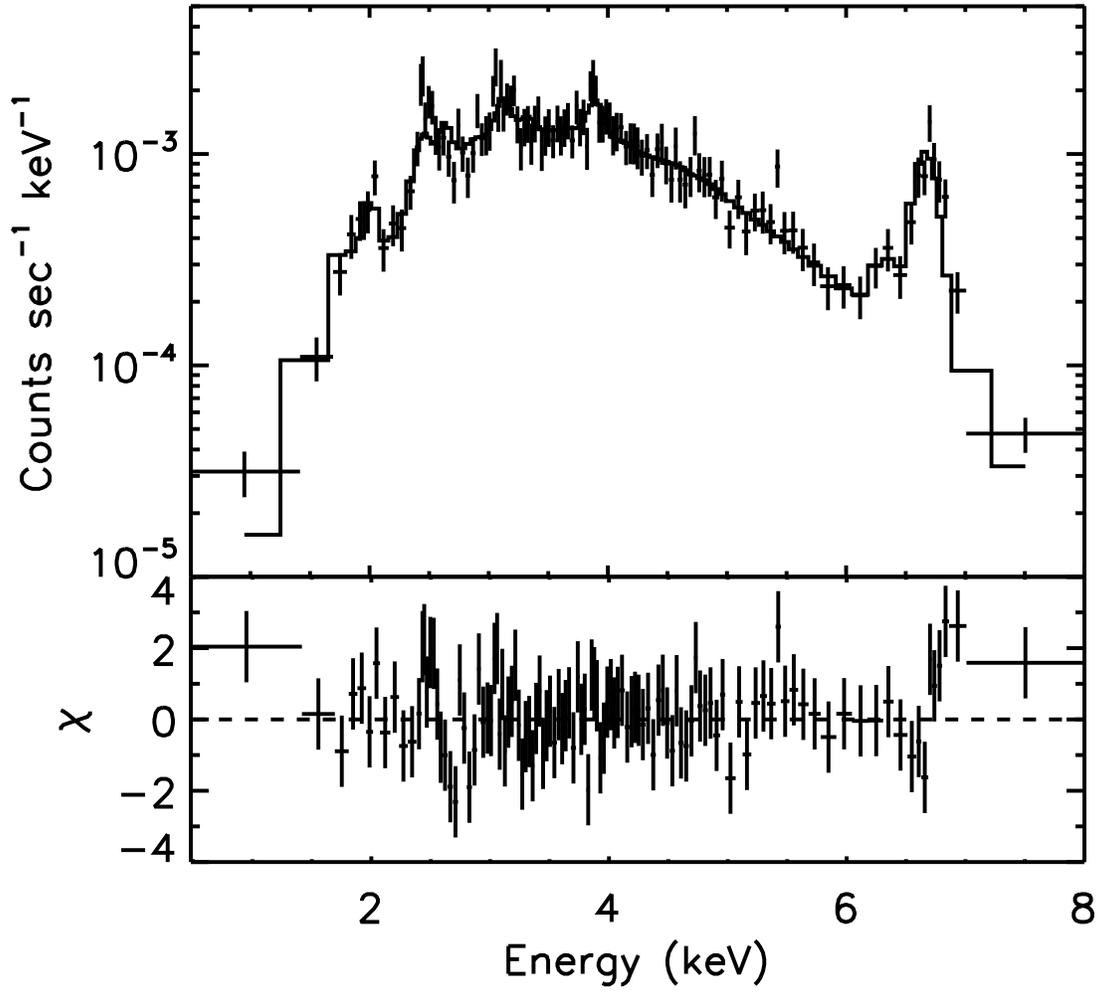}
\caption{The X-ray spectrum of Edd-1. The source displays prominent line emission from the $n$=2--1 transitions of He-like S at 2.4 keV, Ar at 3.1 keV, Ca at 3.9 keV, and Fe at 6.7 keV.}
\label{fig5}
 \end{figure}

The average X-ray spectrum of Edd-1 is displayed in 
Figure~\ref{fig5}. The most prominent feature is line emission
centered at 6.7 keV from the $n$=2--1 transition of He-like Fe with and equivalent width of 2.2keV. In 
addition, lines are evident from S at 2.4 keV, Ar at 3.1 keV, and Ca
at 3.9 keV. Typically accreting binaries are modeled with a multi-temperature blackbody plus a power-law model, however, such a model would not reproduce the prominent line emission observed in Edd-1. The presence of these lines motivates us to model the spectrum 
as a thermal plasma \citep[{\tt apec} in {\tt XSPEC};][]{arn96} 
whose emission has been absorbed by interstellar gas and scattered by interstellar 
dust. We find that the spectrum is consistent with a 
$kT = 2.0^{+0.5}_{-0.2}$ keV thermal
plasma absorbed by a column density of 
$N_{\rm H} = 3.5^{+0.3}_{-0.4} \times 10^{22}$ cm$^{-2}$. Despite the 
prominence of the line emission, the abundances of S, Ar, Ca, and Fe are 
consistent with the solar values (see Table \ref{tbl-3}). 
\begin{deluxetable}{lcc}
\tablecolumns{3}
\tablewidth{0pc}
\tablecaption{{\it Chandra} Spectrum of Edd-1\label{tbl-3}}
\tablehead{
\colhead{Parameter} & \colhead{1--$kT$}  & \colhead{2--$kT$} \\
}
\startdata
$N_{\rm H}$ ($10^{22}$ cm$^{-2}$) & $3.5^{+0.3}_{-0.4}$ &
5.2\tablenotemark{a} \\
$kT_{\rm s}$ (keV) & \nodata & $0.7^{+0.1}_{-0.1}$ \\
$K_{\rm EM, s}$ ($10^{56}$ cm$^{-3}$) & \nodata  & $32^{+9}_{-2}$ \\
$kT_{\rm h}$ (keV) & $2.0^{+0.5}_{-0.2}$ & $4.6^{+0.7}_{-0.7}$ \\
$K_{\rm EM, h}$ ($10^{56}$ cm$^{-3}$) & $5^{+1}_{-1}$ &
$1.4^{+0.4}_{-0.2}$ \\
$Z_{\rm S}/Z_{{\rm S},\odot}$ & $1.1^{+0.5}_{-0.4}$ &
$0.9^{+0.2}_{-0.2}$ \\
$Z_{\rm Ar}/Z_{\rm Ar,\odot}$ & $1.1^{+0.7}_{-0.7}$ &
$1.3^{+0.7}_{-0.4}$ \\
$Z_{\rm Ca}/Z_{\rm Ca,\odot}$ & $1.5^{+0.8}_{-0.8}$ &
$2.2^{+1.3}_{-0.7}$ \\
$Z_{\rm Fe}/Z_{\rm Fe,\odot}$ & $1.4^{+0.3}_{-0.3}$ &
$1.5^{+0.2}_{-0.2}$ \\
$F_{\rm Fe-XXV}$ ($10^{-7}$ photon cm${-2}$ s$^{-1}$) & $3\pm2$
& $2\pm1$ \\
$\chi^2/\nu$ & 125.2/107 & 105.6/105 \\
$F_{\rm X}$ ($10^{-13}$ erg cm$^{-2}$ s$^{-1}$) & $1.2$ & 1.3 \\
$L_{\rm X,s}$ [$D$/8 kpc]$^2$ ($10^{33}$ erg s$^{-1}$) & \nodata &
110 \\
$L_{\rm X,h}$ [$D$/8 kpc]$^2$ ($10^{33}$ erg s$^{-1}$) & $6$ & 3
\enddata
\tablenotetext{a}{Parameter held fixed.}
\tablecomments{$K_{\rm EM}$ is the emission measure for each plasma
component,
$\int n_e n_{\rm H} dV$ [$D$/ 8 kpc]$^{2}$.
Uncertainties are 90\% confidence intervals
($\Delta \chi^2$ = 2.76). Fluxes and luminosities are reported for
the 0.5--8.0 keV band; most of the observed flux is in the 2--8 keV
bandpass.}
\end{deluxetable}

However, the absorption column corresponds to a reddening of only
$A_V = 20$ mag, which is much lower than that inferred from the IR
spectrum, $A_V = (29-33)$. Although it is common for the
column inferred from the X-ray spectrum to be {\it larger} than
would be derived from the IR spectrum (e.g., if the X-rays
are produced by a neutron star embedded in the wind of its companion),
there is no known physical situation in which one would expect the
column of material absorbing the X-rays to be {\it smaller}. Instead,
our model for the X-ray probably under-estimates the amount of flux
produced below 2 keV by Edd-1, which would cause us
to under-estimate the absorption as well. Therefore, we have added a
second, cooler plasma component to our model, and have fixed the
extinction toward the X-ray source at 
$N_{\rm H} = 5.2\times10^{22}$ cm$^{-2}$ (by \citet{pred95}, $A_V = 29$). The spectrum can be adequately
modeled with two plasma components of temperatures $kT_s = 0.7\pm0.1$ keV and
$kT_h = 4.6\pm0.7$ keV, with the cooler component producing $\sim$30\% of
the observed 2--8 keV flux. However, when we deredden the
X-ray spectrum, we find that the additional soft component
produces an enormous amount of flux between 0.5--2 keV, raising the total
inferred luminosity by a factor of $\sim$20 to
$(1.1\pm 0.3)\times10^{35}$ ergs.
This would make Edd-1 either one of the most luminous
known colliding wind binaries, or a moderately bright accreting black
hole or neutron star. Although most of this luminosity will never be
directly observable, we feel that hypothesizing a large amount of unseen X-ray
flux is more physically reasonable than supposing that the X-ray flux
passes through a smaller column of gas than does the much larger flux of
IR photons. Detailed information about both models is listed in Table \ref{tbl-3}.

\section{Discussion}

\begin{figure}
\plotone{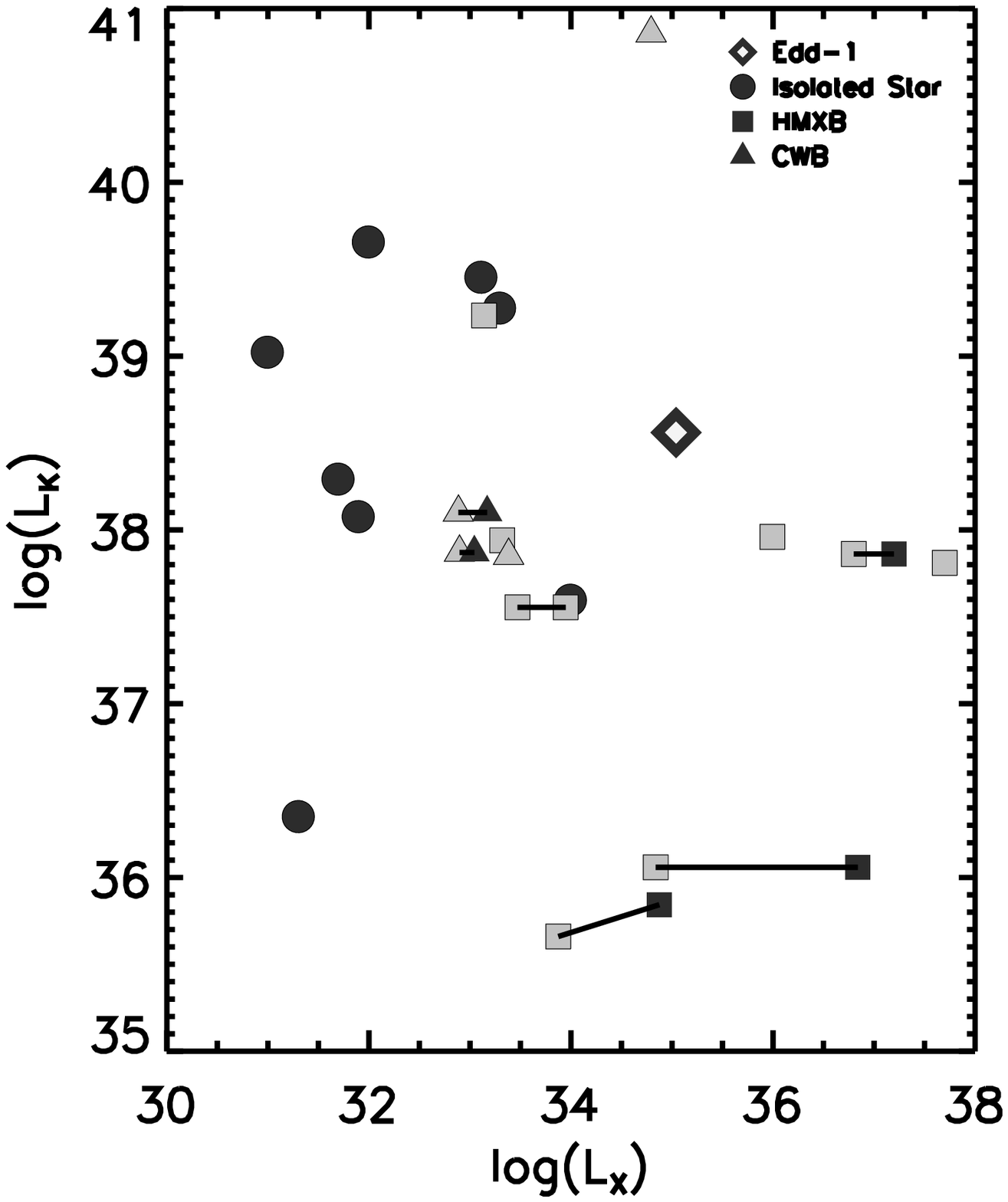}
\caption{Comparison of the X-ray and IR luminosity of Edd-1 to other known X-ray systems containing massive stars. When a single source is observed at varying luminosities, the two points are connected with a line. The lighter symbol indicates a quiescent state. Because simultaneous X-ray and IR data is not typically available, we generally have only one K-band data point. Further information on these sources is available in Table \ref{tbl-4}. }
\label{fig6}
 \end{figure}

In Figure \ref{fig6}, we plot the X-ray and IR luminosity of Edd-1 against known high-mass stars and massive binary systems, including massive OB-stars, Luminous Blue Variables (LBVs), HMXBs, and CWBs. We recognize that the X-ray flux (and the IR to a lesser degree) tends to be variable in binary systems and that our photometric data are not simultaneous. Thus, the specific points are not as useful as the region subtended by the general classes. In addition, there is often uncertainty in both column density and distance for these sources which can shift the individual points. We identify the distances and extinction used in placing these sources as well as the individual source names in Table \ref{tbl-4}.
\begin{deluxetable*}{lllllll}
\tablecaption{{\bf X-ray and IR Source Comparison}}
\tablewidth{0pt}
\tablehead{\colhead{Source} & \colhead{Class} & \colhead{$log(L_X)$\tablenotemark{a}}  & \colhead{$log(L_k)$} & \colhead{d (kpc)} & \colhead{$A_K$} & \colhead{References} }
\startdata
\tableline
\tableline
Edd-1             &  ?         &  35.04    &  38.56   & 8   & 3.4      & [1],[2] \\
\tableline		     	  	  	       	  		   	
CI Cam           &  sgB[e]+X  &  33.54 (q)   &  39.23   & 5	 & 0.3    & [3],[4]\\
\tableline		     	  	  	       	  		   	
IGR J16283-4838  &  Be+NS     &  34 (q)\tablenotemark{b}   &  35.66   & 5	 & 0.9    & [5] \\
IGR J16283-4838  &  Be+NS     &  35 (b)\tablenotemark{b}  &  35.84    & 5	 & 1.4    & [5] \\ 
\tableline		     	  	  	       	  		   	
XTE J1906+090    &  Be+P      &  34.84 (q)   &  36.06    & 4	 & 1.8    &  [6] \\
XTE J1906+090    &  Be+P      &  36.84 (b)   &  36.06    & 4	 & 1.8    &  [6] \\
\tableline		     	  	  	       	  		   	
GRO J2058+42     &  Be+X      &  33.47   &  37.55   & 9.0	 & 1.2    &  [7] \\
GRO J2058+42     &  Be+X      &  33.95   &  37.55   & 9.0	 & 1.2    &  [7] \\
\tableline
X1908+075        &  OBI+NS    &  36\tablenotemark{c}   &  37.96   & 7	 & 2.3    &  [8] \\
\tableline		     	  	  	       	   		   	
Cyg X-1          &  O9I+BH    &  36.80 (q)  &  37.86  & 2.5	 & 0.36    &  [9],[10] \\
Cyg X-1          &  O9I+BH    &  37.20 (b)  &  37.86   & 2.5	 & 0.36    &  [9],[10] \\
\tableline		     	          	       	   		   	
Cen X-3          &  OI+NS     &  37.70\tablenotemark{d}      &  37.81  & 8       & 1.8   & [11],[12] \\  
\tableline
Vela X-1         &  BI+NS     &  33.32      & 37.94   & 1.9     & 0.28  & [13],[14] \\
\tableline
HD 152248        &  O8I+O     &  32.90 (q)   &  37.87     & 1.757	 & 0.15   &  [10],[15],[16] \\
HD 152248        &  O8I+O     &  33.04 (b)   &  37.87     & 1.757	 & 0.15    & [10],[15],[16]  \\
\tableline		     	   	  	       			   	
HD 150136        &  O3+O6V    &  33.38   &  37.85   & 1.32	 & 0.20   &   [10],[17] \\
\tableline		     	  	  	       			   	
$\gamma ^2$ Vel     &  WC8+O7    &  32.89 (q)   &  38.10   & 0.278	 & 0.99    &  [18],[19] \\
$\gamma ^2$ Vel     &  WC8+O7    &  33.17 (b)   &  38.10   & 0.278	 & 0.99    &  [18],[19] \\
\tableline
$\eta$ Car       &  LBV+O	      &  34.88   &  40.86   & 2.3	 & 2.3    &  [22],[23],[24] \\
\tableline		     	          	       	   		   	
P~Cygni          &  LBV	      &  $<$31.0\tablenotemark{e}   &  39.02   & 2.1	 & 0.2	    &  [20],[21] \\
\tableline		     	          	       	   		   	

Pistol           &  LBV	      &  $<$32.0   &  39.66  & 8	 & 3.2	    &  [21],[25] \\
\tableline		     	          	       	   		   	
$\sigma$ Ori E   &  Bp      &  31.30   &  35.94    & 0.40	 & 0.06	    &  [26],[27] \\
\tableline		     	  	  	       	  		   	
HD 108           &  O6f	      &  33.0   &  37.60   & 2.1	 & 0.2	    &  [10],[28],[29] \\
\tableline		     	          	       	   		   	
HD 152408        &  O8Iaf     &  $<$31.7\tablenotemark{e}   &  38.29    & 2.16	 & 0.16    &  [21],[28] \\ 
\tableline		     	          	       	   		   	
HD 151804        &  O8I	      &  31.9\tablenotemark{e}   &  38.07   & 1.66	 & 0.13	    &   [21],[28] \\
\tableline		     	          	       	   		   	
X174516.1        &  O?	      &  33.3   &  39.28   & 8	 & 2.7    &  [21] \\
\tableline		     	          	       	   		   	
H2               &  O?	      &  33.1    &  39.45   & 8	 & 4.5    &  [21] \\
\tableline		     	  	  	       			   	

\enddata
\tablenotetext{a}{Note that X-ray luminosities are reported from a variety of sources, most consistent with a 0.5-8keV range. We further mark objects for which a significantly different energy range is reported as follows (b-e).}
\tablenotetext{b}{Source reports bolometric luminosity based on RXTE observations.}
\tablenotetext{c}{Source reports 1.5-100keV luminosity based on RXTE observations.}
\tablenotetext{d}{Source reports 2-30keV luminosity based on Ginga observations.}
\tablenotetext{e}{Source reports 0.1-2keV luminosity based on ROSAT observations.}
\tablecomments{Identification of sources plotted in Figure \ref{fig6}. Here, we specifiy the source classifications (when available) as well as the distance and K-band extinction used in calculating luminosity. The luminosities are in $erg/s$. The reference numbers are as follows: [1] this work; [2] \citet{mun03}; [3] \citet{boirin02}; [4] \citet{clark00}; [5] \citet{beck05}; [6] \citet{gogus05}; [7] \citet{wilson05}; [8] \citet{morel05}; [9] \citet{schu02}; [10] \citet{maiz04}; [11] \citet{coe97}; [12] \citet{nag92}; [13] \citet{shu02}; [14] \citet{hyl73};[15] \citet{cass80}; [16] \citet{sana04}; [17] \citet{skin05}; [18] \citet{schild04}; [19] \citet{will90}; [20] \citet{turner85}; [21] \citet{muno05}; [22] \citet{vangen94}; [23] \citet{sew01}; [24] \citet{evans03}; [25] \citet{figer98}; [26] \citet{groote82}; [27] \citet{groote04}; [28] \citet{leit84}; [29] \citet{naze04}}
\label{tbl-4}
\end{deluxetable*}

We do not plot Low Mass X-ray Binaries (LMXBs) in Figure \ref{fig6} because LMXBs tend to have $L_X/L_K >>1$. For Edd-1, $L_X/L_K \sim 10^{-4}$.
Even if most of the X-ray emission from Edd-1 is obscured, and its intrinsic luminosity is 100 times larger than what we have inferred, the value of $L_X/L_K$ is more consistent with an HMXB than a LMXB.

In addition to X-ray and IR color, we observe several interesting spectral features which may help identify the nature of this source. Because Edd-1 has P~Cygni profiles in several HeII lines, we have searched in the literature for objects with P~Cygni profiles in HeII in the optical and IR. P~Cygni profiles tend to appear in the HeI lines of HMXBs and CWBs (e.g., Cyg X-1 \citep{gies86}, IGR J16318-4848 \citep{fill04}, and $\eta$ Carinae \citep{hill01}). P~Cygni profiles in HeII lines, such as those observed in Edd-1, are rare. We have also searched for similar X-ray features such as strong Fe-XXV emission in Edd-1. Here, we compare Edd-1 to different types of systems containing massive stars, focusing on similarities to Edd-1's distinguishing spectral features. 

\subsection{Is Edd-1 an Isolated Star?}
We show several OB-stars and LBVs in Figure \ref{fig6} with similar color to Edd-1, including the peculiar Oe-star HD~108. \citet{naze04} observed HD~108 in the optical and X-ray. They observe weak Fe emission in the X-ray at 6.6keV. When fit with a two-temperature plasma, they find $kT_1 \sim 0.2$~keV and $kT_2 \sim$1.4keV, cooler than that observed in Edd-1 when using a two-temperature model, but consistent with the low $A_V$ model. Although some have suggested that HD~108 is a binary because of its strong X-ray emission, long-term observations by \citet{naze04} suggest that HD~108 does not exhibit the same behavior as classical short- or long-term binaries. The H and HeI lines in HD~108 have been observed to change from strong P~Cygni profiles to simple absorptions \citep{naze04}. The fact that Edd-1 shows P~Cygni profiles in HeII, not HeI, suggests the wind in Edd-1 is arising in a hotter region. 
The K-band spectrum of HD~108 shows primarily emission, including Br$\gamma$ and HeI $\lambda 2.114 \mu$m \citep{morris96}, consistent with the IR spectrum of Edd-1. Overall similarities in the X-ray plasma temperatures, infrared emission features, and the presence of a Helium wind suggest that Edd-1 and HD~108 may be similar objects.

The Bp star $\sigma$~Orionis~E, also shares some of Edd-1's distinguishing characteristics. In quiescence, its X-ray thermal temperature is measured between 0.3 and 1.1 keV and reaches 3.5keV in an outburst, consistent with the Edd-1 models. While Fe-XXV (6.7keV) is weakly present, the Fe~K$_\alpha$ (6.4keV) appears in excess during a flare \citep{sanz04}. In contrast, Edd-1 shows weak Fe~K$_\alpha$ emission and strong Fe-XXV emission.  The Helium lines in $\sigma$~Orionis~E vary strongly \citep{groote82}, but this is interpreted as inhomogeneous chemical abundances at the stellar surface \citep{rein00}.

LBVs have similar spectra to B[e] stars and are known for strong, variable Br$\gamma$ lines as well as HI and HeI emission. We plot the positions of the LBV P~Cygni and the LBV candidate Pistol in Figure \ref{fig6}. The X-ray luminosities that we cite for these sources are only upper limits \citep[see][]{muno05}. Bright X-ray emission, such as that observed in Edd-1, would not be expected from an isolated LBV. 

In short, many known isolated LBVs do not show strong X-ray emission. The two isolated OB stars we have considered are both modeled as having a cooler thermal X-ray plasma than Edd-1. In addition, the strong Fe-XXV feature in Edd-1 is observed to be weak in isolated stars. Variable H and He lines are observed in isolated stars and P~Cygni profiles can occur, but are rarely observed in HeII as seen in Edd-1. Thus while we cannot rule out this classification, we believe the isolated star scenario is less likely than a binary nature for Edd-1. 

\subsection{Is Edd-1 A High-Mass X-ray Binary?}
Another possible classification for Edd-1 is that it is an HMXB. Evidence in favor of this includes indications of wind activity in the IR spectrum and strong Fe emission. The HMXBs Vela~X-1 and Cen~X-3 have shown Fe emission - the former in eclipse, the latter out of eclipse \citep{shu02,nag92}. It is rare, however, to find iron lines with equivalent 
widths $>$1 keV. In Cen~X-3, both Fe~$K_\alpha$ (6.40keV) and the Fe-XXV triplet are observed \citep{iaria05}. The equivalent widths of these lines are measured at only a few eV when out of eclipse. In contrast, strong line emission is seen during the eclipses of Vela~X-1, with Fe~$K_\alpha$ equivalent width as large as 1.3 keV in deep eclipse \citep[e.g.][]{choi96, shu02}. Neither of these show dominant Fe-XXV emission.  Both of these HMXBs have a supergiant OB companion. Their optical spectra show Hydrogen and Helium in both absorption and emission associated with the star \citep{mouchet80,dupree80}.

A more rare HMXB companion, a supergiant-B[e] star (sgB[e]), has been observed in CI~Camelopardalis (CI~Cam). While several sgB[e] stars have been observed in the Magellanic Clouds \citep{zick86}, they are rarely observed in the Milky Way. Like Edd-1, CI~Cam has prominent line emission in its IR spectrum, but additionally has forbidden Fe lines. Forbidden lines have not yet been identified in Edd-1. In the X-ray, CI~Cam has a large Fe~$K_\alpha$ line with an equivalent width of 940 eV \citep{fill04}, but it does not exhibit strong Fe-XXV emission. 

Recently IGR~J16318--4848 has been tentatively identified as having a sgB[e] companion. While the X-ray Fe~K$_\alpha$ emission from IGR~J16318--4848 is negligible, a HeI wind has been observed with a velocity of $410 \pm 40$km/s \citep{fill04}. The HeII lines in Edd-1 indicate a much weaker wind with a velocity of $170 \pm 70$ km/s. 

P~Cygni profiles are fairly common among high-mass systems; however, they are more often observed in H or HeI than HeII (e.g. Vela-X-1, IGR J16318-4848). To date, only one HMXB has been observed to have a HeII wind: Cyg~X-1. Cyg~X-1 is a black-hole binary with an O-star companion. It shows weak 
Fe~K$_\alpha$ with an equivalent width of only 16eV. \citet{gies86} found that the P~Cygni profile of HeII~$\lambda 4686 \AA$ in Cyg~X-1 varied with orbital phase and the maximum P~Cygni velocity was $\sim$ 100 km/s, similar to that of Edd-1. The varying P~Cygni profile in Cyg X-1 is modeled as a focused stellar wind whose flux related to the mass transfer rate in the system. 

The presence of Helium wind and the prominent H and He emission in the IR spectrum of HMXBs lends support to this scenario for Edd-1. However, although strong Fe~K$_\alpha$ lines are seen in HMXBs, this feature is rarely the Fe-XXV line observed in Edd-1. Finally, while the inferred X-ray luminosity of Edd-1 is more typical of an HMXB than an isolated star, the IR luminosity is fairly high. So although an HMXB scenario is supported, we cannot make a conclusive classification. 

\subsection{Is Edd-1 A Colliding-Wind Binary?}
Strong wind features may indicate a colliding-wind binary (CWB), which is an association of two massive stars with strong winds. The CWBs HD~152248 and HD~150136 have similar X-ray thermal temperature to Edd-1 in the single-temperature model  \citep{sana04,skin05}. For example, $\gamma ^2$ Velorum (hereafter $\gamma ^2$ Vel), a WC8+O7 binary, has a measured temperature $kT=1.5 keV$ \citep{skin01}. Because the WR star dominates line emission, $\gamma ^2$ Vel appears Helium-rich. In WR+O binaries, weak Brackett series emission may be present, but broad Helium lines dominate emission \citep{skin01,var04} It is possible the He-wind observed in Edd-1 comes from an eclipsed or obscured WR companion. However, because Brackett series emission dominates Edd-1, we find this scenario less likely.

At present, Edd-1 holds the greatest similarity to Eta Carinae. Eta Carinae is an unusually bright X-ray/IR source whose nature is not definitive, but models suggest a CWB containing an LBV+O star pair. The greatest difference between Edd-1 and Eta Carinae is the infrared luminosity. Eta Carinae is intrinsically much brighter in the IR. The similarities between Edd-1 and Eta Carinae are primarily in the X-ray. \citet{viotti04} find that the thermal component of Eta Carinae has a temperature of 5.5 keV with $N_H = 4.8 \times 10^{22} cm^{-2}$, comparable to the two-temperature model of Edd-1. 
In addition, Eta Carinae has a sizeable Fe-XXV line varying between 0.9 and 1.5 keV \citep{viotti04}. This is the only source we have found in the literature with a Fe-XXV line of similar equivalent width to that observed in Edd-1. Typically assumed to be an LBV in a binary system, the X-ray emission of Eta Carinae is modeled as arising from the colliding wind. Detection of P~Cygni profiles in Eta Carinae's optical spectrum is consistent with a CWB classification. \citet{stein04} observed variable HeII emission. It is believed to originate in a dense stellar wind, however the energy supply is still debated \citep{mart05}. Because wind from the LBV in Eta Carinae is mostly cool, the HeII emission is believed to be from the companion, suggested to be an O-star. The presence of strong Fe-XXV emission in Eta Carinae, the similar X-ray thermal temperature to CWBs, and the presence of Helium winds in CWBs lend strong support to Edd-1 being a CWB. 

In summary, Edd-1 shares qualities with a variety of high mass systems including isolated stars, HMXBs, and CWBs. While an isolated star scenario is supported by the IR emission spectrum, it is not consistent with the high X-ray luminosity. Edd-1's X-ray luminosity is fairly common for an HMXB,  and although we observe strong Fe~K$_\alpha$ emission, we do not observe strong Fe-XXV emission in HMXBs. The CWB Eta Carinae does show strong Fe-XXV emission, but in general, CWBs are not as X-ray luminous as Edd-1. Each type of source can show a strong He wind like Edd-1, but the HeII lines in these sources rarely have P~Cygni profiles. The HMXB Cyg X-1 is the only source we have found in the literature with a HeII wind of similar velocity to Edd-1. Based on our observations to date, the evidence favors a CWB system similar to Eta Carinae, with accreting binaries coming in close second. An isolated star is probably the least likely scenario, but cannot be ruled out.   Further distinction of these potential classifications may be derived from either the variability of the source or higher resolution spectroscopy. Until then, the classification of Edd-1 is uncertain, save for the presence of a high mass star in the system.

\section{Conclusions}

The demographics of the stellar population of the Galactic Center are only recently coming to light with the advent of high-resolution X-ray and IR astronomy. The abundance of X-ray sources require careful study at multiple wavelengths in order to identify their nature with similar accuracy to that achieved using optically-based classification systems. From spectroscopy we can distinguish reddened objects from those that are intrinsically red. Edd-1 is a reddened source with an estimated extinction $A_V =29$ mag. We have identified Edd-1 as having prominent emission lines in the X-ray and IR. The HeII lines show P~Cygni profiles consistent with a 170~km/s wind. In addition, Edd-1 has very strong Fe-XXV emission in the X-ray, the line having an equivalent width of 2.2keV.

While it is difficult to positively classify Edd-1 based on the X-ray and IR characteristics observed to date, Edd-1's spectral features indicate the presence of a high mass star. We have compared Edd-1 to OB stars, LBVs, HMXBs, and CWBs -- all of which are types of systems containing a massive star. The X-ray and IR color of Edd-1 is somewhat consistent with each of them; however, the prominent spectral features do not match exactly the characteristics of any of these types. We find Edd-1 is most consistent with the CWB Eta~Carinae. Further study of the variability and spectral features in Edd-1 is necessary to solidify such a classification.

\bigskip
\acknowledgements
This publication makes use of data products from the Two Micron All Sky Survey, which is a joint project of the University of Massachusetts and the Infrared Processing and Analysis Center/California Institute of Technology, funded by the National Aeronautics and Space Administration and the National Science Foundation.

Authors are Visiting Astronomers at the Infrared Telescope Facility, which is operated by the University of Hawaii under Cooperative Agreement no. NCC 5-538 with the National Aeronautics and Space Administration, Office of Space Science, Planetary Astronomy Program.

VJM, SSE, and RMB are supported in part by an NSF Grant (AST-0507547).
VJM is also partially supported by a University of Florida Alumni Fellowship.
MPM is supported through a Hubble Fellowship Grant (program number HST-HF-01164.01-A) from the Space Telescope Science Institute which is operated by the Association of Universities for Research in Astronomy, under NASA contact NAS5-26555. The authors thank the IRTF Director for his assistance, and the IRTF/SpeX team for providing an excellent instrument for observers.


\begin{thebibliography}{}
\bibitem[Arnaud et al.(1996)]{arn96} Arnaud, K.A., 1996, Astronomical Data 
 Analysis Software and Systems V,  eds. Jacoby G. and Barnes J., p17, ASP 
 Conf. Series volume 101.

\bibitem[Beckmann et al.(2005)]{beck05} Beckmann, V. et al. 2005, ApJ, 631, 506.
\bibitem[Boirin et al.(2002)]{boirin02} Boirin, L., Parmar, A. N., Oosterbroek, T., Lumb, D., Orlandini, M., Schartel, N. 2002, A\&A, 394, 205.

\bibitem[Cardelli, Clayton, \& Mathis(1989)]{cardelli89} Cardelli J. A., Clayton G. C., \& Mathis J. S. 1989, ApJ, 345, 245.
\bibitem[Cassinelli et al.(1981)]{cass80} Cassinelli, J. P., Waldron, W. L., Sanders, W. T., Harnden, F. R., Jr., Rosner, R., Vaiana, G. S. 1981, ApJ, 250, 677.

\bibitem[Choi et al.(1996)] {choi96} Choi, C. S., Dotani, R., Day, C. S. R., \& Nagase, F. 1996, ApJ, 471, 447.
\bibitem[Clark et al.(2000)]{clark00} Clark, J. S. et al. 2000, A\&A, 356, 50. 
\bibitem[Coe et al.(1997)]{coe97} Coe, M. J., Buckley, D. A. H., Fabregat, J., Steele, L. A., Still, M. D., Torrejon, J. M. 1997, A\&AS, 126, 237.

\bibitem[Cushing et al.(2004)]{cushing04} Cushing~M.~C., Vacca~W.~D., and Rayner~J.~T. 2004, PASP, 116, 362.
\bibitem[Dupree et al.(1980)]{dupree80} Dupree et al. 1980, ApJ, 238, 969.
\bibitem[Evans et al.(2003)]{evans03} Evans, N. R., Seward, F. D., Krauss, M. I., Isobe, T., Nichols, J., Schlegel, E. M., Wolk, S. J. 2003, ApJ, 589, 509.
\bibitem[Figer, McLean, \& Najarro(1997)]{figer97} Figer, D. F., McLean, I. S., \& Najarro, F.. 1997, ApJ, 486, 420.
\bibitem[Figer et al.(1998)]{figer98} Figer, D. F., Najarro, F., Morris, M., McLean, I. S., Geballe, T. R., Ghez, A. M., Langer, N.  1998, ApJ, 506, 384.
\bibitem[Filliatre \& Chaty(2004)]{fill04} Filliatre~P. \& Chaty~S. 2004, ApJ, 616,469.
\bibitem[Forman et al.(1978)]{for78} Forman, W., Jones, C., Cominsky, L., Julien, P., Murray, S., Peters, G., Tananbaum, H., Giacconi, R. 1978, ApJS, 38, 357.
\bibitem[Giacconi et al.(1972)]{giac72} Giacconi, R., Murray, S., Gursky, H., Kellogg, E., Schreier, E., Tananbaum, H. 1972, ApJ, 178, 281.
\bibitem[Gies \& Bolton(1986)]{gies86} Gies, D. R. \& Bolton, C. T. 1986, ApJ, 304, 389.
\bibitem[G\"{o}\u{g}\"{u}\c{s} et al.(2005)]{gogus05} G\"{o}\u{g}\"{u}\c{s}, E., Patel, S. K., Wilson, C. A., Woods, P. M., Finger, M. H., Kouveliotou, C. 2005, ApJ, 632, 1069.
\bibitem[Groote \& Hunger(1982)]{groote82} Groote, D. \& Hunger, K. 1982, A\&A, 116, 64.
\bibitem[Groote \& Schmitt(2004)]{groote04} Groote, D. \& Schmitt, J. H. M. M. 2004, A\&A, 418, 235.
\bibitem[Gursky(1972)]{gur72} Gursky, H. 1972, ApJ, 175, 141.
\bibitem[Hanson, Conti, \& Ricke(1996)]{hanson96} Hanson, M. M.,  Conti, P. S., \& Ricke, M. J. 1996, ApJ, 107, 281.

\bibitem[Hillier et al.(2001)]{hill01} Hillier, D. J., Davidson, K., Ishibashi, K., Gull, T. 2001, ApJ, 304, 389.

\bibitem[Hyland \& Mould(1973)]{hyl73}Hyland, A. R., \& Mould J. R. 1973, ApJ, 186, 993.

\bibitem[Iaria et al.(2005)]{iaria05} Iaria, R., Di Salvo, T., Robba, N. R., Burderi, L., Lavagetto, G., Riggio, A. 2005, ApJ, 634, 161.
\bibitem[Leitherer \& Wolf(1984)]{leit84} Leitherer, C., Wolf, B. 1984, A\&A, 132, 151.
\bibitem[Ma\'{i}z-Apell\'{a}niz et al.(2004)]{maiz04} Ma\'{i}z-Apell\'{a}niz, J., Walborn, N. R., Galu\'{e}, H. A., Wei, L. H. 2004, ApJS, 151, 103.

\bibitem[Martin et al.(2006)]{mart05} Martin, J. C., Davidson, K., Humphreys, R. M., Hillier, D. J., Ishibashi, K. 2006, ApJ 640, 474. 
\bibitem[McNamara et al.(2000)]{mcn00} McNamara et al. 2000, PASP, 112, 202.
\bibitem[Morel \& Grosdidier(2005)]{morel05} Morel, T., \& Grosdidier, Y. 2005, MNRAS, 356, 665.

\bibitem[Morris, Eenens, \& Blum(1996)]{morris96} Morris, Eenens, \& Blum. 1996, ApJ, 470, 597.
	
\bibitem[Mouchet, Ilovaisky, \& Chevalier (1980)]{mouchet80} Mouchet, M., Ilovaisky, S. A., \& Chevalier, C. 1980, A\&A, 90,113.

\bibitem[Muno et al.(2003)]{mun03} Muno, M. P. et al. 2003, ApJ, 589,225.
\bibitem[Muno et al.(2004)]{mun04} Muno, M. P. et al. 2004, ApJ, 613, 326.
\bibitem[Muno et al.(2004b)]{mun04b} Muno, M. P. et al. 2004b, ApJ, 613, 1179.
\bibitem[Muno et al.(2006)]{muno05} Muno, M. P., Bower, G. C., Burgasser, A. J., Baganoff, F. K., Morris, M. R., Brandt, W. N. 2006, ApJ,638,183
\bibitem[Nagase et al.(1992)]{nag92} Nagase, F., Corbet, R. H. D., Day, C. S. R., Inoue, H., Takeshima, T., Yoshida, K., Mihara, T. 1992, ApJ, 396, 147 

\bibitem[Naz\'{e} et al.(2004)]{naze04}Naz\'{e}, Y., Rauw, G., Vreux, J.-M., \& De Becker, M. 2004, A\&A, 417, 667.


\bibitem[Predehl \& Schmitt(1995)]{pred95} Predehl \& Schmitt. 1995, A\&A, 293, 889.

\bibitem[Rayner et al.(2003)]{rayner03} Rayner~J.~T., Toomey~D.~W., Onaka~P.~M., Denault~A.~J., Stahlberger~W.~E., Vacca~W.~D., Cushing~M.~C., and Wang~S. 2003, PASP, 115, 362.

\bibitem[Reiners et al.(2000)]{rein00} Reiners, A., Stahl, O., Wolf, B., Kaufer, A., Rivinius, T. 2000, A\&A, 363, 585.

\bibitem[Sana et al.(2004)]{sana04} Sana, H., Stevens, I. R., Gosset, E., Rauw, G., \& Vreux, J. M. 2004, MNRAS, 350, 809.
\bibitem[Sanz-Forcada et al.(2004)]{sanz04} Sanz-Forcada, J., Franciosini, E., \& Pallavinci, R. 2004, A\&A, 421, 715S.	

\bibitem[Schild et al.(2004)]{schild04}Schild, H. et al.\ 2004, A\&A, 422, 177.

\bibitem[Schultz et al.(2005)]{schu05} Schultz, A., et al. 2005, "NICMOS Instrument Handbook", Version 8.0, (Baltimore: STScI).

\bibitem[Schulz et al.(2002)]{shu02} Schulz, N. S., Canizares, C. R., Lee, J. C., \& Sako, M. 2002, ApJ, 564, L21.
\bibitem[Schulz et al.(2002b)]{schu02} Schulz, N. S., Cui, W., Canizares, C. R., Marshall, H. L., Lee, J. C., Miller, J. M., Lewin, W. H. G. 2002b, ApJ, 565, 1141.
\bibitem[Seward et al.(2001)]{sew01}Seward, F. D., Butt, Y. M., Karovska, M., Prestwich, A., Schlegel, E. M., \& Corcoran, M. 2001, ApJ, 553, 832.

\bibitem[Skinner et al.(2001)]{skin01}Skinner, S. L., G\"{u}del, M., Schmutz, W., \& Stevens, I. R. 2001, ApJ, 558, L113.
\bibitem[Skinner et al.(2005)]{skin05} Skinner, S. L. Zhekov, S. A., Palla, F., \& Barbosa,  C. L. D. R. 2005, MNRAS, 361, 191.
\bibitem[Steiner \& Damineli(2004)]{stein04}Steiner, J. E. \& Daminel, A. 2004, ApJ, 612, 133
\bibitem[Townsley et al.(2002)]{tow02} Townsley, L. K. et al., 2002, NIM-A, 486, 751
\bibitem[Turner(1985)]{turner85} Turner, D. G. 1985, A\&A, 144, 241.

\bibitem[Vacca et al.(2003)]{vacca03} Vacca~W.~D., Cushing~M.~C. and Rayner~J.~T. 2003, PASP, 115, 389.
\bibitem[van Genderen et al.(1994)]{vangen94} van Genderen, A. M., de Groot, M. J. H., The, P. S. 1994, A\&A, 283, 89.
\bibitem[Varricatt, Williams, \& Ashok(2004)]{var04} Varricatt, W. P., Williams, P. M., Ashok, N. M. 2004, MNRAS, 351, 1307.
\bibitem[Viotti et al.(2004)]{viotti04} Viotti R. F., Antonelli L. A., Rossi C., Rebecchi S. 2004, A\& A, 420, 527. 
\bibitem[Wallace et al.(2000)]{wallace00} Wallace, Meyer, Hinkle, \& Edwards. 2000, ApJ, 535, 325.
\bibitem[Williams et al.(1990)]{will90} Williams, P. M., van der Hucht, K. A., Sandell, G., The, P. S. 1990, MNRAS, 244, 101.
\bibitem[Wilson et al.(2005)]{wilson05} Wilson, C. A., Weisskopf, M. C., Finger, M. H., Coe, M. J., Greiner, J., Reig, P., Papamastorakis, G. 2005, ApJ, 622, 1024.
\bibitem[Zickgraf et al.(1986)]{zick86} Zickgraf, F.-J., Wolf, B., Leitherer, C., Appenzeller, I., Stahl, O. 1986, A\&A, 163,119.
\end{thebibliography}
\end{document}